# Spreadsheets and Long Term Corporate Survival

Grenville J. Croll
EuSpRIG – European Spreadsheet Risks Interest Group
grenvillecroll@gmail.com
v2.012

**ABSTRACT**

*We have conducted an empirical investigation into the long term survival rates of some small but representative samples of the 30,000 largest UK limited companies. These companies were either a control or known to have used, or been interested in the use of, spreadsheets, spreadsheet based monte carlo simulation software, other spreadsheet and decision analysis software and/or related management training. We show that there is a material and statistically significant increase in the long term survival rate of all of these groups of companies compared to the control.*

## 1 INTRODUCTION

The ubiquity of spreadsheets [Croll, 2005] combined with their propensity for error [Panko, 2000] suggests that we should consider whether or not they confer a long term economic advantage to the individuals and companies that use them [Caulkins et al., 2007].

The purpose of this paper is to document an empirical investigation into the long term survival rates of companies known to have used, or been interested in the use of, spreadsheets, spreadsheet based monte carlo simulation software, other spreadsheet and decision analysis software and/or related management training.

We give an overview of related previous work, describe the origin and methods of extraction of the data & controls used in our study, relevant UK company law & company registration processes. We then tabulate, summarize & compare the results and draw conclusions.

## 2 RELATED PREVIOUS WORK

During March 1998, MacMillan [MacMillan, 2000] conducted a series of semi-structured interviews into the use of decision analysis techniques by managers working at anonymous companies involved in the upstream Oil & Gas industry on the UK Continental Shelf (UKCS).

The range of decision analysis techniques explored during the interviews ranged from very simple quantitative techniques such as payback & rate of return, through risk, uncertainty and monte carlo simulation onto advanced methods such as real options, preference and portfolio theory. Zero, one and two points were awarded according to none, some or full use of each technique. Of the 31 companies active on the UKCS, 27 participated in the study, and were ranked in terms of their total usage of the various decision analysis techniques.

Data on corporate performance including Proved Reserves, Market Capitalisation, Total Base Value, Prudential Securitised Risk and Number of Employees were obtained from published sources. The companies were then ranked in order by each of these and several other measures.

A statistical analysis of this ranking data using Spearman's Rank Correlation test showed significant, and sometimes quite strong positive relationships between the scored use of decision analysis techniques and the above listed measures of corporate performance.

The above study is important in terms of the present study for a number of reasons including in particular the fact that the data is contemporaneous to our own data. Furthermore it is near certain that most of the companies and some of the people involved in MacMillan's study were in the population of data from which we have independently drawn our own random samples.

See the Spreadsheet Engineering Research Project (SERP) [Baker et al, 2005a&b] for relatively recent survey work on spreadsheet users including [Powell et al, 2006] which is partly based upon a cohort of users of the Crystal Ball monte carlo simulation product.

**3 TEST DATA ORIGIN**

During the period September 1993 until April 1999, the author was the owner & sole director of Eastern Software Publishing Ltd (ESP) based in Colchester, Essex, UK. ESP purchased the sales database and employed some staff from 4-5-6 World Ltd, a Lotus 1-2-3 spreadsheet add-on specialist also based in Colchester and co-founded by the author in 1986. ESP specialised in the sale of spreadsheet based or spreadsheet related Decision Support Software, other utility software and the provision of related management training and consulting services. The main software products sold, many of which were spreadsheet add-ons for Microsoft Excel, are listed in Table 1.

**Table 1 – ESP Software Products**

| Name | Application | Type |
|---|---|---|
| @Risk for MS Excel | Monte carlo simulation | Excel Addin |
| @Risk for MS Project | Monte carlo simulation | Project Add-in |
| Crystal Ball | Monte carlo simulation | Excel Addin |
| Bestfit | Distribution Fitting | Excel Addin |
| Forecast Pro | Time Series Analysis | Excel File Compatible |
| Smartforecasts | Time Series Analysis | Excel File Compatible |
| What's Best! | Optimisation | Excel Add-in |
| Evolver | Genetic Algorithms | Excel Add-in |
| Neuframe | Neural Networks | Windows App |
| Simul8 | Process Simulation | Windows App |
| EasyFlow | Flowcharting | Windows App |
| BPS | Financial Planning | Excel Spreadsheet |
| BPS | Activity Based Costing | Excel Spreadsheet |
| DPL | Decision Analysis | Windows App |
| Daisy | Data Visualisation | Excel File Compatible |
| XpertRule | Knowledge Management | Windows App |
| Visual Baler | Spreadsheet Compiler | Lotus 1-2-3 V4 Compiler |
| Spreadsheet Professional | Spreadsheet Auditing | Excel File Compatible |
| Monarch | Spreadsheet Data Conversion | Excel File Compatible |
| Bond & Options Analyst | Function libraries | Excel Functions |
| Utilities | Various | Windows App |

Regular Management seminars were held on Risk Analysis [Croll, 1995], Business Forecasting, Linear Programming, Database scoring, Neural Networks and Accounting for Non-Accountants. Approximately 1,500 delegates attended over 150 public and private management seminars held mainly in London.

More than half of ESP's turnover was of the @Risk monte carlo simulation add-in tool for Microsoft Excel & Project and related "Introduction to Risk Analysis" and "Risk Management for Large Projects" management seminars.

ESP was essentially a marketing company which used a combination of catalogue based direct mail, fax broadcasting and outbound telemarketing to achieve its sales. The company distributed 50,000 software catalogues, 500,000 faxes and made over 55,000 outbound telephone calls per annum on average. The company employed 10 staff with a turnover of in excess of £500,000pa in each of 1994-1998. The average sale price of products and seminars sold was £350. ESP ceased trading when the UK distribution contract for @Risk was transferred to another UK company in April 1999.

The data for this study is a complete copy of the ESP sales database, DBFEB99 comprising over 60,000 records. This database was exported from the networked CRM system in February 1999. Each record contains the name, job title, company name, company address and summary sales information for each enquirer, prospect and customer of ESP. There is an associated marketing database HIST0299 which recorded the summary outcome of every inbound and outbound telephone conversation which comprises 374,000 records which was used in computing the database scores.

A substantial proportion of people on the ESP database were recruited through a post carding campaign. Several million pre-paid business reply postcards with the tag lines "Get the Most out of Lotus", "Get the Most out of Excel" or "Get the Spreadsheet Software Catalogue" were distributed in a variety of UK & European computer and business media in the early 1990's. The database was organically developed through the ensuing years by the 4-5-6 World and ESP sales staff. It is safe to assume that all the people on the ESP database were spreadsheet users.

**4 CONTROL DATA ORIGIN**

During 1996 ESP wished to expand its sales activities in order to address a wider audience. ESP obtained a license for the use of a database of the 50,000 Largest UK Companies from a reputable information provider. A de-duplication was performed in order to identify those companies – approximately 20,000 - that ESP did not already hold information on. This new data was scored by the scoring system described below, prior to its upload to the ESP CRM system and the commencement of substantive marketing activities by telephone, fax and catalogue. Very few sales were made and these activities were rapidly discontinued, however the data has remained available and comprises the control data (CONTROL) for this study.

**5 DATABASE SCORING SYSTEM**

Given the size of the ESP database and the relatively small staff, a database scoring system was essential in order to prioritize the people and companies to whom sales and marketing attention should be devoted. The scoring system used approximately 16 non financial & non sales database variables to compute a value between zero and one which represented the likelihood that a database record (corresponding to a person) was a customer. A customer was denoted by a one and a non-customer by a zero. The variables typically used in the scoring system are given in Table 2.

**Table 2 ESP Database Scoring Variables**

| | |
|---|---|
| Next Call Date | Date of the next scheduled telephone call date |
| Date Entered | Date entered on the database |
| Senior Manager | *if* the job title was director, manager, senior etc |
| Enquirer | *if* an inbound enquiry had been made |
| Source | *if* the data source was known |
| Eire | *if* the client was located in Eire |
| Europe | *if* the client was located in Europe |
| Dealer | *if* the client was a software re-seller |
| High Mail | *if* the client was a regular catalogue recipient |
| Positive Call | # +ve call outcomes (interested / very interested) |
| Negative Call | # -ve call outcomes (not interested / do not call) |
| Incoming Call | # inbound calls |
| Correspondence | # outbound letters sent |
| Employees | # employees |

Binary variables are denoted by *if*. Date and # variables are positive integers.

The scoring was performed through a multiple regression in Lotus 1-2-3 version 2.4 on a random 12.5% sample of the ESP database (abut 8,000 records). The R-squared of the regression was typically around 27% and the t's of the variables ranged from 2 to 25, indicating that the statistical strength of the scoring model was extremely high (*F>150*) and all of the above variables were good or very good contributors to this overall strength. Once the scores had been computed and checked on the sample they were computed for the whole database and the control database using the coefficients of the sample. From time to time an independent second sample was obtained and regressed in order to check the regression coefficients. Neural network techniques were tried and abandoned as they were less predictive than simple least squares.

The scoring system was exceptionally effective in directing sales efforts and achieving sales - the *a posteori* effectiveness of the scoring system was regularly reviewed against achieved sales. During searches of the database, use of records with scores greater than 0.5 identified mostly customers, with a few hot prospects. Use of a score greater than 0.1 identified a wider selection of prospects to whom marketing materials could be profitably sent. Records with a score less than 0.05 were generally ignored. Staff, whose remuneration included a commission element, soon became used to using the higher scoring records to direct their outbound telemarketing activities. The database was re-scored approximately quarterly so that new and revised sales data was scored appropriately.

## 6 UK COMPANY REGISTRATION

In order to gain the advantage of limited liability, UK entrepreneurs can form limited liability companies through which they can trade. The company becomes legally separate from the shareholders who own the company. The shareholders can appoint themselves or others as directors in order to manage the company's affairs. Any debts of the company are (ordinarily) separate from the affairs of the shareholders or directors as individuals.

The incorporation and dissolution of limited companies is recorded in the UK by companies house ( www.companieshouse.co.uk ). Companies House is the government operated registration body through which incorporation, annual financial reporting, dissolution, changes of company names, directors, shareholders and addresses etc takes place. At incorporation every limited company, which must have a unique name, is allocated a unique company number. Smaller companies are denoted by the words

Limited or Ltd occurring after the name. Larger companies are often, but not always identified by the words Public Limited Company or PLC after the name.

Companies House offers a free Web Check service by which means anybody can inspect the basic records pertaining to a UK limited company. We have made extensive use of this free service in the present study in order to determine the names, company numbers, company name changes and dates of dissolution of UK companies.

This analysis has been exclusively confined to UK Limited and Public Limited Companies contained on the ESP database and the control. Unincorporated bodies and government institutions in particular, of which there were many, could not be included.

**7 PRELIMINARY STUDY**

**7.1 Outline & Objective**

Given the ubiquity, importance and criticality of spreadsheets within companies and given the well understood high error rate of spreadsheets, we wanted to see if variations in spreadsheet use within UK companies were correlated with variations in company survival rates as recorded at Companies House.

As we had no absolutely no idea about rates of long-term corporate survival or dissolution, we arbitrarily decided to take a random sample of 20 people $n$ who had attended the authors "Introduction to Risk Analysis" seminar [Croll, 1995] during 1993-1996, the period of ESP's trading, and compare that with 20 people $m$ who had bought other products and services from ESP during that period.

We would then check the names of the 40 companies for whom these people worked against Companies House to determine the company number. We knew we would not be able to get a perfect match every time and so $n_1$ & $m_1$ would be the numbers of people for whom we had an exact or near match of the company name. Once we had identified a company name and corresponding company number, we could determine via Companies House Web Check service if the company was still trading at the date of this study – Jan 2012. The numbers of companies still trading would be $n_2$ & $m_2$.

We would thus have two survival rates $s_1$ & $s_2$ where:

$$s_1 = n_2 / n_1$$
$$s_2 = m_2 / m_1$$

In order to be able to determine if any difference between $s_1$ and $s_2$ was statistically significant, we would use the Test of Two Proportions, which due to binomial considerations requires that:

$$n_1 \ s_1 \quad > 5$$
$$n_1 \ (1-s_1) \quad > 5$$
$$m_1 \ s_2 \quad > 5$$
$$m_1 \ (1-s_2) \quad > 5$$

Thus the specific, written and *a priori* objective of our preliminary study was:

> *"was there any difference in the long term survival rate of ESP customers 1993-1999 between those who bought management training ('Introduction to Risk Analysis') and those who did not"*

In the event, we performed two independent preliminary experiments as outlined above on Sample A and Sample B so that we had four survival rates: $s_{1A}, s_{2A}$ and $s_{1B}, s_{2B}$

## 7.2 Data Acquisition

We were surprised to find that it was relatively easy to match a company name on the ESP database with a corresponding entry at Companies House. Despite the common changes of names of companies and the incorporation of companies with similar or even identical names (particularly around times of corporate distress), the Companies House database enabled us to find, on almost every occasion, the company name and number (i.e. the legal identity) of the company with whom ESP was almost certainly trading at the time. This required careful interrogation of the three main Companies House databases: Current company names; Dissolved company names and Previous company names. Having found the correct company number, determining its present trading status was trivial as this information was immediately displayed. Companies were either Active, Active (Dormant), Dissolved, Liquidated or about to be Struck Off. Since ESP must have traded with companies that were Active, all other company status codes reflected a greater or lesser degree of subsequent corporate failure.

This study was indifferent to the reasons behind each company dissolution, whether it be due to voluntary or compulsory insolvency, corporate acquisition or whatever. Our only concern was whether the company with whom ESP had initially traded was still Active or not, as recorded on the Companies House database.

When interrogating the companies house database with a company name from the ESP database, we had to make sure that a) the company had been incorporated prior to the end of the 1993-1999 period of our investigation and b) the company had not been dissolved prior to 1993. There was a little subjectivity in the matching of company names to entries on the Companies House Database. Most often the exact legal identity of the company was obvious, but occasionally we had to take into consideration small typographic changes & errors, common abbreviations and slight changes in punctuation. We determined the match of the ESP company name to the Companies House company name to be either Exact, Near or None. When a company had been dissolved we recorded the date of dissolution if it was available. If the match of the company name was not exact, we recorded a note giving details and recorded the Match status as Near. Where there was no match against a name on Companies House, the Match status was recorded as None and that company was excluded from all further analysis.

We then recorded in a spreadsheet database the outcome of our interrogation of the Companies House database for preliminary samples A and B.

## 7.3 Preliminary Results

By way of illustration, we reproduce the data for our preliminary study in Appendix A. From this data the survival rates for Samples A and B are immediately apparent:

**Sample A – Risk Analysis Seminar Attendees**

$$s_{1A} = n_{2A} / n_{1A} \quad = 13 / 20 \quad = \quad 65\%$$

**Sample A – Other Customers**

$$s_{2A} = m_{2A} / m_{1A} \quad = 16 / 18 \quad = \quad 88\%$$

**Sample B – Risk Analysis Seminar Attendees**

$$s_{1B} = n_{2B} / n_{1B} \quad = 16 / 20 \quad = \quad 80\%$$

**Sample B – Other Customers**

$$s_{2B} = m_{2B} / m_{1B} \quad = 16 / 20 \quad = \quad 80\%$$

Unfortunately due to the Binomial consideration outlined above we were not able to use the Test of Two proportions introduced above to determine if the difference between the two survival rates in each sample were statistically significant. However since both samples were independent, we could combine the data for both samples into a Combined Sample, Sample C.

**Sample C – Risk Analysis Seminar Attendees**

$$s_{1C} = n_{2C} / n_{1C} \quad = 29 / 40 \quad = \quad 72.5\%$$

**Sample C – Other Customers**

$$s_{2C} = m_{2C} / m_{1C} \quad = 32 / 38 \quad = \quad 84.2\%$$

This time we were able to use the test of two proportions to determine that there was no statistically significant difference between the survival rates in the two samples.

**8 MAIN STUDY**

**8.1 Determining Sample size**

Although the lack of statistical significance in the preliminary study was disappointing, we were encouraged by the very high Exact and Near match rate of 78 / 80 = 97.5%. Plus we had an order of magnitude for corporate survival in the range 72%-84% over a period of approximately 16 years – the midpoint of the ESP data being 1996 and the present date being 2012. We were also able to estimate the amount of time that would be required to interrogate the Companies House database as it proved possible to match and retrieve survival and dissolution data at a rate of about 30 companies per hour.

We were thus able to plan out a main study. In order to be able to detect a statistically significant difference between two proportions of 72.5% and 84% at *p<0.1* i.e. 90% statistical significance, the sample size would have to be at least 68.

**8.1 Determining Database Segments to Investigate**

We decided to increase the number of differing types of ESP data to investigate. There were myriad ways to dice and slice the ESP database. The segments we chose reflected the fact that half of the business was due to @Risk related products and seminars, that other customers were a prime target for ongoing sales activities, and that enquirers were less likely to be sales targets. Following MacMillan, we ranked the segments in order of the perceived relative use of Decision Analysis software and methods, with 6 being the highest use.

**Table 3A – ESP & Control Database Segments, Descriptions & Rank**

| SEGMENT KEY | SEGMENT DESCRIPTION | SEGMENT RANK |
|---|---|---|
| CONTROL | Large UK Companies Control | 1 |
| ENQ | ESP Enquirers – anyone asking for a catalogue or information on a specific product or service who was not already a customer | 2 |
| OTHER | Purchaser of a product other than @Risk | 3 |
| RISK | Purchaser of the @Risk Monte Carlo Simulation Excel Spreadsheet add-in | 4 |
| SEM | Attendee at the "Introduction to Risk Analysis" Management Seminar | 5 |
| RISK&SEM | Purchaser of @Risk Monte Carlo Simulation Excel Spreadsheet add-in and attendee at the "Introduction to Risk Analysis" Management Seminar | 6 |

We also report results for a combined RISK, SEM and RISK&SEM segment which is denoted ALLRISK and is useful where sample sizes are reduced. The synthetic ALLRISK segment is omitted from the ranking results.

**8.2 Database Segment Detail**

The ESP Database segments and the control represent a continuum of spreadsheet use represented by the Segment Rank.

**8.2.1 Control**

At the lowest end were members of the Control segment. These people had never responded to any direct marketing, telemarketing, or other advertising material distributed by ESP or 4-5-6 World by virtue of the de-duplication performed. Other than the fact that the data was contemporaneous to the ESP data and complementary to it by virtue of it being the other half of the UK's largest 50,000 UK companies at the time, little else specific was known about this data. No doubt they used spreadsheets, however we have no information on the level of such usage and as a result of this we have placed them at the lowest segment ranking in terms of our *a priori* perception of their spreadsheet usage. To have placed them anywhere else in the segment ranking would have been perverse.

**8.2.2 Enq**

One step up from the Control data were the ESP enquirers. These people had made a positive effort to get in touch with ESP by responding to ESP's spreadsheet and decision science related marketing efforts by mail, phone, fax or email. ESP held basic information about them including their name, company name, address, phone number, fax number and email address. ESP also held coded information about the products or services they had enquired about and the media source of their enquiry. ESP Enquirers may have subsequently bought software or services promoted by ESP elsewhere. It is certain that members of the ENQ segment were spreadsheet users.

### 8.2.3 Other

At the next stage were purchasers of products *other than* the @Risk monte carlo Excel spreadsheet add-in and/or related management training. Clearly, purchasing spreadsheet add-ons and spreadsheet related decision science products indicates an increased interest and intensity in the use of spreadsheets and related decision science technology. Included within the OTHER category are purchasers of Crystal Ball, the @Risk competitor product (Crystal Ball sales volume was substantially lower than the @Risk sales volume) and attendees at other management seminars which were generally related to the spreadsheet based software products which ESP sold. This category includes the purchase of some utility software which was not spreadsheet related, however this proportion was a relatively small percentage of ESP's turnover by value.

### 8.2.4 Risk

We placed purchasers of the @Risk Excel spreadsheet add-in at the next level of the segment ranking. @Risk was and is a complex software product which is deeply bound into the spreadsheet paradigm. In order to understand how the product operates, even on a basic level, requires intimate knowledge of a spreadsheet's design, function, purpose and features. Command of the @Risk product requires sound knowledge of a number of statistical principles (mean, median, mode, skewness, kurtosis, distributions of various types, sampling, graphics) and the monte-carlo computational method itself.

### 8.2.5 Sem

We placed attendees at the author's "Introduction to Risk Analysis" management seminar at the penultimate segment ranking. This was a one day tutorial which covered the principles of quantitative and qualitative risk management, the specific topics listed in the previous section, and provided an opportunity to use the @Risk for Excel product during the course of several group exercises. The seminar was held monthly at an Hotel in London or on-site at company premises.

There was a choice between giving RISK and SEM a tied ranking, RISK the higher ranking or SEM the higher ranking. We chose the latter as the seminar placed the @Risk product in context, enabling its use to be delegated as a specialism.

### 8.2.6 Risk & Sem

We gave people who purchased the @Risk software product and attended the Introduction to Risk Analysis" Management Seminar the highest Segment Ranking.

### 8.3 Data Dates & Ages

All ESP data used in this (and the preliminary) study had a "Date_Updated" field with a value between 1$^{st}$ Jan 1993 and 31$^{st}$ December 1999. This indicated that an ESP staff member had checked the correctness of that record on the date indicated in the Date_Updated field by virtue of a phone call made or received, a letter or email sent or received or a sale made or returned. Note that the date of availability of the control data was November 1996 – midway through ESP's lifetime and the lifetime of the rest of the data used.

**8.4 Sampling the ESP database**

We set up a database query to randomly sample the above segments of the ESP database in order to produce approximately 70 records per segment and 420 records total.

We decided to perform our experiment twice on two independently drawn random samples from the ESP data. The first random sample had 424 records and the second had 427 records. Since the results from the two samples were remarkably similar, we report only the combined results of the two samples amounting to 851 records.

The ESP database sometimes contained the details of many individuals working at the same company. We therefore decided to de-duplicate the data in each of our two samples such that within each sample there would only be one individual from each company, with no regard paid as to which individuals records from the same company should be retained or deleted. This data is labelled COMPANIES in the results reports. The original data containing data for individuals is identified as INDIVIDUALS in the results reports.

Larger companies might be more inclined or able to buy software and services or have a greater propensity for survival, and so we have separately tabulated the data for the generally larger PLC's and the generally smaller LIMITED COMPANIES.

Having obtained results for the first two samples, we decided to investigate the ESP database further by generating a third sample comprising people who worked in the City of London (Postcodes beginning EC1, EC2, EC3, EC4 & E14) [Croll, 2005]. This data is denoted CITY OF LONDON INDIVIDUALS in the results reports. We again de-duplicated this sample to produce a set of results for CITY OF LONDON COMPANIES. Data volume for the City of London sample was lower than that for the first two samples.

We document the population and sample sizes in for the two original samples and the City of London sample in Tables 3B & 3C below.

**Table 3B – ESP Database & Control Segments, Population & Sample Sizes – First Two Samples**

| SEGMENT NAME | TOTAL DATA | % TOTAL | LTD & PLCS 1993-1999 | % | SAMPLE SIZE | % |
|---|---|---|---|---|---|---|
| CONTROL | 19,271 | 31.8% | 18,134 | 54.0% | 147 | 0.8% |
| ENQ | 33,723 | 55.6% | 12,583 | 37.5% | 146 | 1.2% |
| OTHER | 6,159 | 10.2% | 2,320 | 6.9% | 140 | 6.0% |
| RISK | 809 | 1.3% | 264 | 0.8% | 140 | 53.0% |
| SEM | 372 | 0.6% | 138 | 0.4% | 138 | 100.0% |
| RISK&SEM | 321 | 0.5% | 140 | 0.4% | 140 | 100.0% |
| ALLRISK | 1,502 | 2.5% | 542 | 1.6% | 418 | 77.1% |
| TOTAL | 60,655 | 100.0% | 33,579 | 100.0% | 851 | 2.5% |

**Table 3C – ESP Database & Control Segments, Population & Sample Sizes – City of London Sample**

| SEGMENT NAME | TOTAL DATA | % TOTAL | LTD & PLCS 1993-1999 | % | SAMPLE SIZE | % |
|---|---|---|---|---|---|---|
| CONTROL | 1,144 | 35.3% | 1,056 | 62.4% | 67 | 6.3% |
| ENQ | 1,598 | 49.3% | 469 | 27.7% | 66 | 14.1% |
| OTHER | 356 | 11.0% | 125 | 7.4% | 65 | 52.0% |
| RISK | 80 | 2.5% | 15 | 0.9% | 15 | 100.0% |
| SEM | 39 | 1.2% | 12 | 0.7% | 12 | 100.0% |
| RISK&SEM | 22 | 0.7% | 15 | 0.9% | 15 | 100.0% |
| ALLRISK | 141 | 4.4% | 42 | 2.5% | 42 | 100.0% |
| TOTAL | 3,239 | 100.0% | 1,692 | 100.0% | 240 | 14.2% |

**8.5 Experimental method**.

We used an identical methodology to that described in the Preliminary Study to extract information on corporate survival from the Companies House database. In addition, we extracted and recorded the Database Score for each individual and tabulated an average database score per segment DSCORE in the results.

**8.6 Results**

**8.6.1 Survival Rates**

We show in Table 4 the percentage of companies that remained active in each ESP database segment since trading with ESP until the present day – the long term corporate survival rate. The colour coding shows the statistical significance of the differences between the survival rates for each ESP segment and that for the CONTROL segment in each of the six result categories. Table 5 shows the statistical significance represented by the colours used.

**Table 4 – Main Study Long Term Corporate Survival Rates**

**SUMMARY RESULTS**

**ESP ACTIVE% VS CONTROL ACTIVE%**
**TEST OF TWO PROPORTIONS**

| SEGNAME | INDIVIDUALS | COMPANIES | LIMITED COMPANIES | PUBLIC LIMITED COMPANIES | CITY OF LONDON INDIVIDUALS | CITY OF LONDON COMPANIES |
|---|---|---|---|---|---|---|
| CONTROL | 54.4% | 54.4% | 54.5% | 53.3% | 55.2% | 54.5% |
| ENQ | 60.6% | 61.5% | 60.0% | 68.0% | 59.7% | 55.8% |
| OTHER | 61.7% | 62.1% | 58.6% | 92.3% | 60.7% | 61.2% |
| RISK | 68.6% | 69.1% | 62.5% | 92.6% | 80.0% | 66.7% |
| SEM | 71.2% | 73.0% | 70.5% | 82.6% | 100.0% | 100.0% |
| RISK&SEM | 69.9% | 71.3% | 66.2% | 87.0% | 80.0% | 85.7% |
| ALLRISK | 69.9% | 71.0% | 66.3% | 87.7% | 85.7% | 80.8% |

**Table 5 – Colour coding for Statistical Significance tests**

| | |
|---|---|
| *99% Significance* | 🟧 |
| *95% Significance* | 🟧 |
| *90% Significance* | 🟨 |

The detailed data behind the above summary showing the number of records per segment, the match counts for exact, near and no match, average database scores per record, net active and ACTIVE% are given in Tables B1-B6 of Appendix B. The full set of proportion tests from which the above summary statistics are derived are given in Tables C1-C6 of Appendix C.

**8.7 Survival Rate Rankings & Correlations**

For each of the six categories of data, we used Simple Least Squares and Spearman's method to investigate the Linear and Rank correlations (expressed as $R^2$) between 1) the ACTIVE% and the average database scores per record DSCORE and 2) the ACTIVE% and the ESP Segment Rank. These four correlations are reported in Table 6. The colour coding indicates the statistical significance of the correlations. The raw data for these correlations is given in Tables B1-B6 of Appendix B.

**Table 6 – Main Study Long Term Corporate Survival Rate Correlations**

SUMMARY CORRELATIONS (R^2)
DEP VAR IS ACTIVE %

| INDEP VAR | METHOD | INDIVIDUALS | COMPANIES | LIMITED COMPANIES | PUBLIC LIMITED COMPANIES | CITY OF LONDON INDIVIDUALS | CITY OF LONDON COMPANIES |
|---|---|---|---|---|---|---|---|
| DSCORE | SPEARMAN | 94% | 83% | 77% | 71% | 64% | 60% |
| DSCORE | LINEAR | 78% | 75% | 60% | 62% | | |
| RANK | SPEARMAN | 94% | 94% | 89% | 54% | 90% | 94% |
| RANK | LINEAR | 83% | 82% | 63% | 53% | 57% | 60% |

**9 SUMMARY OF RESULTS**

The first set of results are summarised in Table 4. The survival rates within each of the following three clusters of results are completely independent of each other and are in addition directly comparable:

    Cluster 1)    Limited Companies, Public Limited Companies, City of London Companies
    Cluster 2)    Individuals and City of London Individuals
    Cluster 3)    Companies and City of London Companies

A number of exogenous factors which may have a bearing on survival rates have not been included in this study. These factors include company size, employees, turnover, profitability, industrial sector and the acquisition and subsequent dissolution of successful and profitable companies as a result of their success. Note that when a company is acquired by another UK Limited company, the acquirer remains within the population being studied. We have purposefully investigated survival rates in the City of London geographic area [Croll, 2005][Croll, 2009].

The results in Table 4 clearly show that the survival rates in the control segments were the same in all three clusters and across all six groups of individuals and companies studied. This suggests that no significant exogenous factors have been omitted from this study. Thus any changes in survival rate must be attributable only to the spreadsheet related variables under consideration.

The results in Table 4 show a statistically significant positive increase in survival rates between Companies and Individuals who were purchasers of either or both of the @Risk software and related management training compared with the control. Over a period of 16 years, approximately 55% of the limited companies in the control segment survived whereas between 68% and 73% of the Individuals and Limited Companies in the @Risk and/or seminar segments survived *(p<0.05*).

These effects were still apparent when the data was split between smaller Limited Companies and generally larger PLC's. The survival rate in the PLC's AllRisk sample was very high at 87.7% and contrasted most sharply *(p<0.01)* with the lower survival rate in the PLC control segment. The very high survival level (>80%) compared to the control persisted in the City of London AllRisk sample at both an individual and a company level *(p<0.01 & p<0.05)*. The very high survival levels in the PLC's & City of London AllRisk samples are significantly higher than the high survival levels seen in the Individuals, Companies and Limited Companies AllRisk samples *(p<0.05, see Table C7 in Appendix C)*.

There was very little difference in the survival rates across the Control, Enquirer and Other segments. This suggests that although these samples were quite small, they were representative of their much larger populations. Within these three segments, the City of London groups are no different from the wider geographic population of companies.

There was very little difference in the survival rates between results based on companies or results based on individuals within companies. Thus conclusions could be drawn from results for individuals where the sample size for companies is too small.

The results in Table 6 show that there is a gradual and near monotonic increase in survival rate from the control segment, through enquirers, other customers, then purchasers of one, other or both of @Risk and the related management training in most of the six groups of companies studied. The data is both rank correlated and linearly correlated not only with the *a priori* segment ranking based on MacMillan's study, but also with the database scores which were historically used to quantitatively rank customer, enquiry and control data.

The summary and correlation results in Tables 4 & 6 show that there is a small positive difference in survival rates between the Enquirer & Other segments where spreadsheet use is certain and the control segment where the use of spreadsheets & decision science software was unknown. Over the 16 years of this study whereas about 55% of the control segments survived about 60% of the Enquirer & Other segments survived. The Control segment comprises 54% and the Enquirer and Other segments comprise 44.4% of the total population of Limited companies studied.

Less than 2% of the 30,000 largest companies in the UK exhibited the significantly higher long term survival rate which has been correlated with the use of monte carlo simulation software and related management training during the period of this study.

## 10 CONCLUSION

We have conducted an empirical investigation into the long term survival rates of some small but representative samples of the 30,000 largest UK limited companies. These companies were either a control or known to have used, or been interested in the use of, spreadsheets, spreadsheet based monte carlo simulation software, other spreadsheet and decision analysis software and/or related management training. We show that there is a material and statistically significant increase in the long term survival rate of all of these groups of companies compared to the control. These results are consistent with the earlier results of [MacMillan, 2000] who used contemporaneous data.

For a small proportion of spreadsheet users, the use of spreadsheet based monte carlo simulation software and related management training is correlated with a significantly increased long term corporate survival rate. This effect is particularly pronounced within PLC's and companies located within the City of London.

## ACKNOWLEDGEMENTS

We thank the staff of ESP for their skill in creating and maintaining such a fine information asset as the ESP database. We thank the ESP seminar delegates for teaching us more than we taught them. We thank Peter Harrison and the anonymous referee for their detailed, informative and constructive comments. We thank Fiona MacMillan and the EuSpRIG 2012 delegates for their valued feedback.

## REFERENCES

Baker, K.R., Powell, S., & Lawson, B., (2005a) "Spreadsheet Engineering Research Project", http://mba.tuck.dartmouth.edu/spreadsheet  15:07 20/5/05

Baker, K.R., Foster-Johnson, L., Lawson, B., Powell, S.G., (2005b) "A Survey of MBA Spreadsheet Users" Spreadsheet Engineering Research Project, Accessed 5th April 2009, 15:00 http://mba.tuck.dartmouth.edu/spreadsheet/product_pubs_files/SurveyPaper.doc

Powell, S.G., Lawson, B., Foster-Johnson, L., (2006), "Comparison of Characteristics and Practices amongst Spreadsheet Users with Different Levels of Experience", Proc. European Spreadsheet Risks Interest Group, http://arxiv.org/abs/0803.0168

Caulkins, P., Morrison, E.L., Weideman, T., (2007) "Spreadsheet Errors and Decision Making: Evidence from Field Interviews", Journal of Organizational and End User Computing, Volume 19, Issue 3, http://www.infoscionline.com/downloadPDF/pdf/ITJ3782_UQOaYcOa45.pdf
Accessed 17th March 2009 17:17

Croll, G.J., (1995), Cost and Schedule Risk Analysis in Major Engineering Projects, IEE Colloquium on Future Developments in Project Management Systems 1995/183, London, UK

Croll, G.J. (2005) "The Importance and Criticality of Spreadsheets in the City of London", Proc. European Spreadsheet Risks Interest Group, http://arxiv.org/abs/0709.4063

Croll, G.J., (2009), Spreadsheets and the Financial Collapse, Proc. European Spreadsheet Risks Int. Grp.(EuSpRIG)  145-161 http://arxiv.org/abs/0908.4420

MacMillan, F. (2000) "Risk, Uncertainty and Investment Decision-Making in the Upstream Oil and Gas Industry",  Ph.D. Thesis, University of Aberdeen, Scotland.

Panko, (2000) "Spreadsheet Errors: What We Know. What We Think We Can Do", European Spreadsheet Risks Interest Group, 1st Annual Symposium, University of Greenwich, pp7-18.
http://arxiv.org/abs/0802.3457

# APPENDIX A - PRELIMINARY STUDY RAW DATA

## SAMPLE A - Seminar Attendees

| Company | Match | Status | Dissolution Date | Company Number | Notes |
|---|---|---|---|---|---|
| CRE Group Ltd | Exact | Dissolved | 09/10/2010 | 2924220 | |
| Norweb Plc | Exact | Dissolved | 08/10/1996 | 2300085 | |
| Parsons Group int. Ltd | Exact | Active | | 3393325 | |
| Nortel Plc | Exact | Liquidation | | 2515751 | |
| HVR Consulting Services Ltd | Exact | Proposal to Strike off | | 1775338 | |
| British Aerospace Plc | Exact | Active | | 1470151 | |
| Marley Building Materials Ltd | Exact | Dissolved | 27/03/2009 | 263226 | |
| Texaco Ltd | Exact | Active | | 145197 | |
| British Nuclear Fuels Plc | Exact | Active | | 1002607 | |
| National Westminster Bank Plc | Exact | Active | | 929027 | |
| Neotronics Ltd | Exact | Liquidation | | 2046915 | |
| WRC Plc | Exact | Active | | 2262098 | |
| Barclays Bank Plc | Exact | Active | | 48839 | |
| M W Kellogg Ltd | Exact | Active | | 909986 | |
| Texaco Ltd | Exact | Active | | 145197 | |
| Amec Process & Energy Ltd | Near | Dissolved | 12/5/2009 | 2797300 | Note 1 |
| Barclays Bank Plc. | Exact | Active | | 48839 | |
| Powergen Plc | Exact | Active | | 2366970 | |
| Rand Information Systems Ltd | Exact | Active | | 1217993 | |
| N M T Group Plc | Exact | Active | | SC170841 | |
| **Totals** | **20** | **13 Active** | | | |

## SAMPLE A - Other Customers

| Company | Match | Status | Dissolution Date | Company Number | Notes |
|---|---|---|---|---|---|
| Drumgrange Ltd | Exact | Active | | 1460044 | |
| The Tennant Rubber Company Ltd | Near | Active | | 548218 | Note 2 |
| Taurus Training Services Ltd | Exact | Active | | 2493688 | |
| Planco Consulting Ltd | Near | Active | | 2908077 | Note 3 |
| British Gas Plc | Exact | Active | | 2006000 | |
| Pfh Computer Systems (Cork)Ltd | None | None | | | Note 4 |
| Nashblend Ltd T/A Westcountry Tiling | Exact | Active | | 2840842 | |
| Mech-mail Envelopes Ltd | Exact | Active | | 2413935 | |
| British Alcan Lynemouth Ltd | Exact | Dissolved | 08/11/2011 | 644321 | |
| The Co-operative Bank Plc. | Exact | Active | | 990937 | |
| How Group Plc | Exact | Active | | 1984855 | |
| WAM (GB) Ltd. | Exact | Active | | 1868307 | |
| Argo Software Ltd | Exact | Active | | 1820200 | |
| Pilkington Communications Ltd | Near | Active | | 522707 | Note 5 |
| E T P M DeepSea Ltd | Exact | Active | | 1902584 | |
| Perkins Slade Ltd | Exact | Active | | 969374 | |
| Lawson Mardon (Mi) Ltd | Exact | Liquidation | | 186479 | |
| Nat West Bank Plc | None | None | | | Note 6 |
| Allied Colloids Ltd | Exact | Active | | 722043 | |
| Mitsubishi Corporation Finance Plc | Exact | Active | | 1865061 | |
| **Totals** | **18** | **16 Active** | | | |

Note 1: Amec Process and Energy Construction Ltd
Note 2: Tennant Rubber Company (The) Ltd
Note 3: Planco consulting (uk) Ltd
Note 4: No UK Match, No Irish Match
Note 5: Pilkington Communication Systems Ltd
Note 6: Too many possible names

**SAMPLE B - Seminar Attendees**

| Company | Match | Status | Dissolution Date | Company Number | Notes |
|---|---|---|---|---|---|
| Parsons Group Int. Ltd | Exact | Active | | 3393325 | |
| Cable & Wireless Plc | Exact | Active | | 238525 | |
| Costain Oil, Gas & Process Ltd | Exact | Active | | 786418 | |
| Scottish Hydro Electric Plc | Exact | Active | | SC117119 | |
| Texaco Ltd | Exact | Active | | 145197 | |
| Barclays Bank Plc | Exact | Active | | 48839 | |
| Stolt Comex Seeley UK Ltd | Near | Active | | 974791 | Note 7 |
| Gec Marconi Dynamics Ltd | Exact | Dissolved | 13/03/2007 | 622657 | |
| Royal Doulton Plc | Exact | Liquidation | | 452813 | |
| ICI Chemicals & Polymers Ltd | Exact | Active | | 358535 | |
| QBE International Insurance Ltd | Exact | Active | | AC001566 | |
| Babcock & Brown Ltd | Exact | Liquidation | | 2645511 | |
| CRE Group Ltd | Exact | Dissolved | 09/10/2010 | 2924220 | |
| Project Management International Plc | Exact | Active | | 2100456 | |
| British Telecom Plc | Exact | Active | | 1800000 | |
| Nat Air Traffic Services Ltd | Near | Active | | 3155567 | Note 8 |
| Provident Personal Credit Ltd | Exact | Active | | 146091 | |
| Landis & Gyr Ltd | Near | Active | | 1202284 | |
| Lloyds Bank Plc | Exact | Active | | 2065 | |
| Marshall Of Cambridge Aerospace Ltd | Exact | Active | | 245740 | |
| **Totals** | **20** | **16 Active** | | | |

**SAMPLE B - Other Customers**

| Company | Match | Status | Dissolution Date | Company Number | Notes |
|---|---|---|---|---|---|
| Henderson Hardware Ltd | Exact | Active | | 4522572 | |
| Leopold Joseph & Sons Ltd | Exact | Active | | 338594 | |
| Waddington Galleries Ltd | Exact | Active | | 872520 | |
| Kerry Group Plc | Exact | Active | | 6547046 | |
| ICI Chemicals & Polymers Ltd | Exact | Active | | 358535 | |
| P.P.G Industries Uk Ltd. | Exact | Active | | 2110620 | |
| Kurt Mueller (Uk) Ltd | Exact | Active | | 477895 | |
| Devonport Management Ltd | Exact | Active | | 2052982 | |
| Acorn Computers Ltd/ | Exact | Active | | 1403810 | |
| Lee James Electronics Ltd. | Exact | Active | | 1575997 | |
| Technigold Comp Services Ltd | Exact | Dissolved | 13/06/2000 | 2231027 | |
| Vax Ltd | Exact | Active | | 1341840 | |
| GKN Defence Ltd | Exact | Active | | 617410 | |
| Charles Day Steels Ltd | Exact | Active | | 1289020 | |
| I G E Medical Systems Ltd | Exact | Active | | 252567 | |
| Visionhire Ltd | Exact | Dissolved | 27/12/2011 | 473581 | |
| Phosyn plc | Exact | Active | | 1035807 | |
| Executive Computers (UK) Ltd | Exact | Dissolved | 08/10/2002 | 3562064 | |
| Cad-Capture Ltd | Exact | Liquidation | | 1786020 | |
| The Robert McBridge Group Ltd | Near | Active | | 220175 | Note 9 |
| **Totals** | **20** | **16 Active** | | | |

Note 7: Stolt Comex Seaway Ltd
Note 8: National Air Traffic Services Ltd
Note 9: Robert Mcbride Group Ltd (The)

# APPENDIX B – MAIN STUDY – DETAILED RESULTS

## Table B1 – Individuals

**INDIVIDUALS**

| SEGNAME | SEGRANK | DSCORE | COUNT | NEAR | NONE | NET | ACTIVE | ACTIVE% |
|---|---|---|---|---|---|---|---|---|
| CONTROL | 1 | 0.00 | 147 | 5 | 0 | 147 | 80 | 54.42% |
| ENQ | 2 | 0.05 | 146 | 6 | 9 | 137 | 83 | 60.58% |
| OTHER | 3 | 0.14 | 140 | 11 | 12 | 128 | 79 | 61.72% |
| RISK | 4 | 0.28 | 140 | 17 | 3 | 137 | 94 | 68.61% |
| SEM | 5 | 0.28 | 138 | 12 | 6 | 132 | 94 | 71.21% |
| RISK&SEM | 6 | 0.43 | 140 | 8 | 4 | 136 | 95 | 69.85% |
| | | | | | | | | |
| ALLRISK | | 0.33 | 418 | 37 | 13 | 405 | 283 | 69.88% |
| | | | | | | | | |
| TOTAL | | | 851 | 59 | 34 | 817 | 525 | 64.26% |

## Table B2 - Companies

**COMPANIES**

| SEGNAME | SEGRANK | DSCORE | COUNT | NEAR | NONE | NET | ACTIVE | ACTIVE% |
|---|---|---|---|---|---|---|---|---|
| CONTROL | 1 | 0.00 | 147 | 5 | 0 | 147 | 80 | 54.42% |
| ENQ | 2 | 0.05 | 144 | 6 | 9 | 135 | 83 | 61.48% |
| OTHER | 3 | 0.15 | 136 | 10 | 12 | 124 | 77 | 62.10% |
| RISK | 4 | 0.28 | 126 | 17 | 3 | 123 | 85 | 69.11% |
| SEM | 5 | 0.27 | 117 | 11 | 6 | 111 | 81 | 72.97% |
| RISK&SEM | 6 | 0.45 | 96 | 5 | 2 | 94 | 67 | 71.28% |
| | | | | | | | | |
| ALLRISK | | 0.33 | 339 | 33 | 11 | 328 | 233 | 71.04% |
| | | | | | | | | |
| TOTAL | | | 766 | 54 | 32 | 734 | 473 | 64.44% |

## Table B3 – Limited Companies

**LIMITED COMPANIES**

| SEGNAME | SEGRANK | DSCORE | COUNT | NEAR | NONE | NET | ACTIVE | ACTIVE% |
|---|---|---|---|---|---|---|---|---|
| CONTROL | 1 | 0.00 | 132 | 4 | 0 | 132 | 72 | 54.55% |
| ENQ | 2 | 0.05 | 118 | 4 | 8 | 110 | 66 | 60.00% |
| OTHER | 3 | 0.14 | 122 | 9 | 11 | 111 | 65 | 58.56% |
| RISK | 4 | 0.28 | 99 | 13 | 3 | 96 | 60 | 62.50% |
| SEM | 5 | 0.28 | 94 | 10 | 6 | 88 | 62 | 70.45% |
| RISK&SEM | 6 | 0.47 | 73 | 5 | 2 | 71 | 47 | 66.20% |
| | | | | | | | | |
| ALLRISK | | 0.33 | 266 | 28 | 11 | 255 | 169 | 66.27% |
| | | | | | | | | |
| TOTAL | | | 638 | 45 | 30 | 608 | 372 | 61.18% |

**Table B4 - Public Limited Companies**

**PUBLIC LIMITED COMPANIES**

| SEGNAME | SEGRANK | DSCORE | COUNT | NEAR | NONE | NET | ACTIVE | ACTIVE% |
|---|---|---|---|---|---|---|---|---|
| CONTROL | 1 | -0.01 | 15 | 1 | 0 | 15 | 8 | 53.33% |
| ENQ | 2 | 0.07 | 26 | 2 | 1 | 25 | 17 | 68.00% |
| OTHER | 3 | 0.19 | 14 | 1 | 1 | 13 | 12 | 92.31% |
| RISK | 4 | 0.26 | 27 | 4 | 0 | 27 | 25 | 92.59% |
| SEM | 5 | 0.25 | 23 | 1 | 0 | 23 | 19 | 82.61% |
| RISK&SEM | 6 | 0.38 | 23 | 0 | 0 | 23 | 20 | 86.96% |
| | | | | | | | | |
| ALLRISK | | 0.30 | 73 | 5 | 0 | 73 | 64 | 87.67% |
| | | | | | | | | |
| TOTAL | | | 128 | 9 | 2 | 126 | 101 | 80.16% |

**Table B5 - City of London - Individuals**

**CITY OF LONDON INDIVIDUALS**

| SEGNAME | SEGRANK | DSCORE | COUNT | NEAR | NONE | NET | ACTIVE | ACTIVE% |
|---|---|---|---|---|---|---|---|---|
| CONTROL | 1 | -0.01 | 67 | 3 | 0 | 67 | 37 | 55.22% |
| ENQ | 2 | 0.05 | 66 | 3 | 4 | 62 | 37 | 59.68% |
| OTHER | 3 | 0.15 | 65 | 5 | 4 | 61 | 37 | 60.66% |
| RISK | 4 | 0.29 | 15 | 3 | 0 | 15 | 12 | 80.00% |
| SEM | 5 | 0.14 | 12 | 2 | 0 | 12 | 12 | 100.00% |
| RISK&SEM | 6 | 0.31 | 15 | 0 | 0 | 15 | 12 | 80.00% |
| | | | | | | | | |
| ALLRISK | | 0.26 | 42 | 5 | 0 | 42 | 36 | 85.71% |
| | | | | | | | | |
| TOTAL | | | 240 | 16 | 8 | 232 | 147 | 63.36% |

**Table B6 – City of London – Companies**

**CITY OF LONDON COMPANIES**

| SEGNAME | SEGRANK | DSCORE | COUNT | NEAR | NONE | NET | ACTIVE | ACTIVE% |
|---|---|---|---|---|---|---|---|---|
| CONTROL | 1 | -0.01 | 66 | 3 | 0 | 66 | 36 | 54.55% |
| ENQ | 2 | 0.05 | 56 | 3 | 4 | 52 | 29 | 55.77% |
| OTHER | 3 | 0.17 | 53 | 4 | 4 | 49 | 30 | 61.22% |
| RISK | 4 | 0.31 | 12 | 3 | 0 | 12 | 8 | 66.67% |
| SEM | 5 | 0.11 | 7 | 1 | 0 | 7 | 7 | 100.00% |
| RISK&SEM | 6 | 0.31 | 7 | 0 | 0 | 7 | 6 | 85.71% |
| | | | | | | | | |
| ALLRISK | | 0.26 | 26 | 4 | 0 | 26 | 21 | 80.77% |
| | | | | | | | | |
| TOTAL | | | 201 | 14 | 8 | 193 | 116 | 60.10% |

## APPENDIX C – MAIN STUDY – RELATIVE SURVIVAL RATES

Using the Active% column of each of Tables B1-B6 above, we calculated the ratio between the survival rates of each segment. We then performed a test of two proportions on the survival rate ratio and indicated its statistical significance according to the simple colour code of Table 5. For example in Table B1 above, the survival rate for Risk purchasers & Seminar delegates was 69.85% whereas the survival rate for the Control was 54.42%. The ratio between these two survival rates is 69.85 / 54.42 =  1.28 which ratio was significant at the 99% level and is colour coded Dark Orange. Light Orange indicates 95% significance and Yellow indicates 90% significance. In Table C7 we compare the survival rates within the AllRisk segment.

### Table C1 - Individuals

| SEGNAME | CONTROL | ENQ | OTHER | RISK | SEM | RISK&SEM | ALLRISK |
|---|---|---|---|---|---|---|---|
| CONTROL | 100% | 111% | 113% | 126% | 131% | 128% | 128% |
| ENQ | | 100% | 102% | 113% | 118% | 115% | 115% |
| OTHER | | | 100% | 111% | 115% | 113% | 113% |
| RISK | | | | 100% | 104% | 102% | 102% |
| SEM | | | | | 100% | 98% | 98% |
| RISK&SEM | | | | | | 100% | 100% |
| ALLRISK | | | | | | | 100% |

### Table C2 – Companies

| SEGNAME | CONTROL | ENQ | OTHER | RISK | SEM | RISK&SEM | ALLRISK |
|---|---|---|---|---|---|---|---|
| CONTROL | 100% | 113% | 114% | 127% | 134% | 131% | 131% |
| ENQ | | 100% | 101% | 112% | 119% | 116% | 116% |
| OTHER | | | 100% | 111% | 118% | 115% | 114% |
| RISK | | | | 100% | 106% | 103% | 103% |
| SEM | | | | | 100% | 98% | 97% |
| RISK&SEM | | | | | | 100% | 100% |
| ALLRISK | | | | | | | 100% |

### Table C3 – Limited Companies

| SEGNAME | CONTROL | ENQ | OTHER | RISK | SEM | RISK&SEM | ALLRISK |
|---|---|---|---|---|---|---|---|
| CONTROL | 100% | 110% | 107% | 115% | 129% | 121% | 122% |
| ENQ | | 100% | 98% | 104% | 117% | 110% | 110% |
| OTHER | | | 100% | 107% | 120% | 113% | 113% |
| RISK | | | | 100% | 113% | 106% | 106% |
| SEM | | | | | 100% | 94% | 94% |
| RISK&SEM | | | | | | 100% | 100% |
| ALLRISK | | | | | | | 100% |

### Table C4 - Public Limited Companies

| SEGNAME | CONTROL | ENQ | OTHER | RISK | SEM | RISK&SEM | ALLRISK |
|---|---|---|---|---|---|---|---|
| CONTROL | 100% | 128% | 173% | 174% | 155% | 163% | 164% |
| ENQ | | 100% | 136% | 136% | 121% | 128% | 129% |
| OTHER | | | 100% | 100% | 89% | 94% | 95% |
| RISK | | | | 100% | 89% | 94% | 95% |
| SEM | | | | | 100% | 105% | 106% |
| RISK&SEM | | | | | | 100% | 101% |
| ALLRISK | | | | | | | 100% |

### Table C5 - City of London – Individuals

| SEGNAME | CONTROL | ENQ | OTHER | RISK | SEM | RISK&SEM | ALLRISK |
|---|---|---|---|---|---|---|---|
| CONTROL | 100% | 108% | 110% | 145% | 181% | 145% | 155% |
| ENQ | | 100% | 102% | 134% | 168% | 134% | 144% |
| OTHER | | | 100% | 132% | 165% | 132% | 141% |
| RISK | | | | 100% | 125% | 100% | 107% |
| SEM | | | | | 100% | 80% | 86% |
| RISK&SEM | | | | | | 100% | 107% |
| ALLRISK | | | | | | | 100% |

### Table C6 – City of London – Companies

| SEGNAME | CONTROL | ENQ | OTHER | RISK | SEM | RISK&SEM | ALLRISK |
|---|---|---|---|---|---|---|---|
| CONTROL | 100% | 102% | 112% | 122% | 183% | 157% | 148% |
| ENQ | | 100% | 110% | 120% | 179% | 154% | 145% |
| OTHER | | | 100% | 109% | 163% | 140% | 132% |
| RISK | | | | 100% | 150% | 129% | 121% |
| SEM | | | | | 100% | 86% | 81% |
| RISK&SEM | | | | | | 100% | 94% |
| ALLRISK | | | | | | | 100% |

### Table C7 – AllRisk segment

| ALLRISK | INDIVIDUALS | COMPANIES | LIMITED | PUBLIC | CITY IND | CITY COMP |
|---|---|---|---|---|---|---|
| INDIVIDUALS | 100% | 102% | 95% | 125% | 123% | 116% |
| COMPANIES | | 100% | 93% | 123% | 121% | 114% |
| LIMITED | | | 100% | 132% | 129% | 122% |
| PUBLIC | | | | 100% | 98% | 92% |
| CITY IND | | | | | 100% | 94% |
| CITY COMP | | | | | | 100% |